\documentstyle[psfig,prl,aps,twocolumn]{revtex}
\begin{document}
\draft
\title{Equilibrium random-field Ising critical scattering
in the antiferromagnet $\mathbf Fe_{0.93}Zn_{0.07}F_2$}
\author{Z. Slani\v{c} and D. P. Belanger}
\address{Department of Physics, University of California,
Santa Cruz, CA 95064 USA}
\author{J. A. Fernandez-Baca}
\address{Solid State Division, Oak Ridge National Laboratory,
Oak Ridge, TN 37831-6393 USA}
\date{\today}
\maketitle
\begin{abstract}
It has long been believed that equilibrium random-field Ising model (RFIM)
critical scattering studies are not feasible in dilute antiferromagnets
close to and below $T_c(H)$ because of severe non-equilibrium effects.
The high magnetic concentration Ising antiferromagnet $Fe_{0.93}Zn_{0.07}F_2$,
however, does provide equilibrium behavior.  We have employed scaling techniques
to extract the universal equilibrium scattering line shape,
critical exponents $\nu = 0.87\pm0.07$ and $\eta = 0.20 \pm 0.05$, and
amplitude ratios of this RFIM system.
\end{abstract}

Models for the statistical physics of phase transitions can
be experimentally tested by measuring the universal critical
parameters and comparing them with theoretical predictions.
This is crucial for the understanding of difficult problems such as
the random-field Ising model (RFIM).
Although it is well known that a phase transition
occurs\cite{by91} for the $d=3$ RFIM,
the characterization of the equilibrium critical behavior in its
experimental realization, anisotropic randomly dilute
antiferromagnets such as $Fe_xZn_{1-x}F_2$ in applied uniform
fields\cite{fa79}, has not been possible near to and below the transition,
$T_c(H)$, despite years of intense experimental investigation.
This is primarily a result of the severe hysteresis in the scattering
line shapes near to and below $T_c(H)$.
Although long-range order is observed
after cooling in zero field (ZFC) and heating through the transition,
upon cooling through the transition in the field (FC) long-range
antiferromagnetic order is never achieved, even at low
temperatures.  In either case, the line shapes are difficult to
interpret.  It was widely thought for many years that these nonequilibrium effects
were unavoidable.

Since random-field effects increase with the dilution,
as well as the strength of the applied magnetic field\cite{c84},
most early studies focused on magnetic concentrations $x<0.8$
in order to achieve, for reasonable $H$, a suitable range of reduced
temperature, $t=(T-T_c(H))/T_c(H)$, over which asymptotic critical
behavior might be observed.  However, nonequilibrium effects
exist below the equilibrium temperature\cite{kjbr85}
$T_{eq}(H)$ which lies just above $T_c(H)$.  
Much effort was directed to understanding the behavior
of the domains, both near to and below $T_c(H)$.
Neutron scattering studies of the Bragg intensity in the
bulk crystal\cite{bkjn87} $Fe_{0.46}Zn_{0.54}F_2$ and the
epitaxial film\cite{bwshnlrl95}
of $Fe_{0.52}Zn_{0.48}F_2$ showed anomalous behavior even under ZFC
caused by domain formation.  It was shown that, at low $T$, RFIM dynamics
are dominated by domains\cite{hbkn92,lsboh93}.
The dominance of domain effects near $T_c(H)$ was evident in ac
susceptibility measurements\cite{bkk95}.  From these studies over
several years along with Monte-Carlo (MC) simulations\cite{snu98,hb92}
it became clear that the severe nonequilibrium effects were primarily
due to the large number of vacancies that are well connected and
thus allow the formation of domain walls with insignificant
energy cost.

Recently, we demonstrated\cite{sbf98} that the
metastability problem can, in fact, be overcome
by employing a magnetically concentrated crystal,
$Fe_{0.93}Zn_{0.07}F_2$, in which the domain walls must break
many magnetic bonds to form since the vacancies are well separated\cite{rc98}.
As a consequence, domains do not form under either FC or ZFC
and the line shapes from both processes are found to be identical for all $T$
and, hence, represent equilibrium behavior.  Of course,
the small dilution necessitates the use of large fields
to probe the RFIM critical regime, so we
used the highest field available with the present
apparatus, $H=7$~T.
Attempts to analyze the equilibrium critical scattering in
a preliminary study\cite{sbf98} failed because the sample
had a concentration gradient\cite{specific} that limited
analysis to $|t|> 10^{-3}$.  Hence, we were only able to
attempt an analysis using mean-field (MF) line shapes that
yield values for the inverse correlation, $\kappa$, and 
staggered susceptibility, $\chi_s$, that
do not, however, follow power-law behavior in $|t|$ below $T_c(H)$.
We have overcome these
difficulties by using a thin crystal of thickness $0.44$~mm,
less than $1/10$ the original thickness, cut with its plane
perpendicular to the concentration gradient.
This has made it possible to extend the scattering
measurements to $|t|>10^{-4}$, greatly increasing the
accuracy of the critical behavior analysis and,
by making a few simple assumptions about the scaling properties
of the scattering line shape as described below, we have been able
to go beyond the inaccurate MF line shapes traditionally
used in this type of analysis.  With the use of the scaling techniques
and the additional data from the thin crystal,
we have finally achieved a consistent analysis of the critical
behavior above {\em and below} $T_c(H)$ for the RFIM.

Neutron scattering measurements were made
at the Oak Ridge National Laboratory
High Flux Isotope Reactor using a two-axis spectrometer
configuration.  We used the (0 0 2) reflection
of pyrolytic graphite (PG) at an energy of $14.7$ meV
to monochromate the beam.  We mainly employed two different
collimation configurations.  The lower resolution,
primarily used for the large sample,
is with 70 min of arc before
the monochromator, 20 before the sample and
20 after the sample.  Primarily for the thin sample,
we made scans with 10 min of arc before and after the sample.
PG filters were used to eliminate
higher-order scattering.  The carbon thermometry scale was calibrated
to agree with recent specific heat results \cite{sb97} for the
$H=0$ transition.  The field dependence of the thermometry was also
calibrated.  All scans used in this report are transverse about
the (1 0 0) antiferromagnetic Bragg point.

The observed line shape is given by $S(q)$ convoluted with the
instrumental resolution.  In MF\cite{by91}
\begin{equation}
S(q)= \frac{A} {q ^ 2 + \kappa ^ 2 }
+  \frac{B} {(q ^ 2 + \kappa ^ 2 )^2 } + {M_s}^2 \delta (q) \quad ,
\label{lor_lor2}
\end{equation}
where $\kappa = {\kappa _o}^{\pm}|t|^{\nu}$,
$+$ and $-$ are for $t>0$ and $t<0$,
respectively and $M_s$ is the staggered magnetization.
The first term represents $\chi_s (q)$
and the second two the disconnected susceptibility, ${\chi_s}^{dis}(q)$.
In principle, the amplitudes $A$ and $B$ are constant in $T$ and the same
above and below $T_c(H)$, although in practice this is not true since
real systems do not follow MF behavior.
For translationally invariant systems,
$B=0$.  For real systems MF is inadequate
since we must have the scaling behavior
$\chi_s (q) \propto \kappa ^{\eta -2} f(q/\kappa )$ with the limits
$\chi_s (q) \propto \kappa ^{\eta -2} /(1+q^2/\kappa ^2)$ for
$|q| << \kappa $ and $\chi_s (q) \propto q ^{\eta -2}$ for
$|q| >> \kappa $.  For the $d=3$
pure transition, $\eta \approx 0.03$.  Hence, the MF Lorentzian
line shape is adequate except very close to $T_c(H)$, as
explicitly shown\cite{by87} for $FeF_2$
where the corrections to MF are important for $|t|<10^{-3}$,
particularly for $T<T_c(H)$.  For smaller $|t|$, the $t<0$ Tarko-Fisher\cite{tf75} (TF),
\begin{equation}
f(q/\kappa ) \propto \frac
{(1+\phi ^{\prime \,2}q^{2}/\kappa^{2})^{\sigma+\eta/2}}
{(1+\psi ^\prime  q^{2}/\kappa^{2})(1+\phi ^{\prime \prime\, 2}q^{2}/
\kappa^{2})^{\sigma}}
\mbox{ ,}
\label{eqn:neut-tf}
\end{equation}
and the $t>0$ Fisher-Burford\cite{fb67} (FB),
\begin{equation}
f(q/\kappa ) \propto \frac
{(1+\phi^{2}q^{2}/\kappa^{2})^{\eta/2}}
{1+\psi q^{2}/\kappa^{2}}
\mbox{ ,}
\label{eqn:neut-fb}
\end{equation}
approximants, where $\phi$, $\phi ^{\prime}$, $\phi^{\prime \prime}$,
and $\sigma$ are fixed, and $\psi = 1 + 1/2 \eta \phi ^2$ and
$\psi ^{\prime}=1+1/2 \eta {\phi ^{\prime}}^2 +\sigma ({\phi ^{\prime}}^2-
{\phi ^{\prime \prime}}^2)$,
were found to be excellent for the 
analysis and yielded exponents
in superb agreement with theory\cite{by91}.

The dilute case with $H=0$ corresponds to the random-exchange
Ising model for which the value of $\eta$ is
similarly small and the MF line shape proved adequate
for $|t|>10^{-3}$.  For $x=0.46$, the scattering results\cite{bkj86}
$\nu = 0.69 \pm 0.03$ and $\gamma = 1.33 \pm 0.02$ agree remarkably
well with recent MC results\cite{bfmspr98}
$\nu = 0.684 \pm 0.005$ and $\gamma = 1.34 \pm 0.01$.

In a previous scattering work\cite{sbf98}, we found two important results
using Eq.\ 1 to fit the data for $Fe_{0.93}Zn_{0.07}F_2$ for
$H=7$~T.  First, the results for $\kappa$
and $\chi_s$ are independent of whether the FC or ZFC
procedures were used, i.e.\ the line shapes are
equilibrium ones at all $T$.  Second, the values for $\kappa$
and $\chi_s$ obtained from the line shape analysis could be
fit to a power law in $|t|$ for $T>T_c(H)$,
but not $T<T_c(H)$.  For $T>T_c(H)$,
$\nu = 0.90 \pm 0.01$, $\gamma = 1.72 \pm 0.02$ and
$\bar{\gamma} = 3.0 \pm 0.1 $ were obtained.
The lack of power law behavior for $\kappa$ and
$\chi _s$ below $T_c(H)$ indicates that the MF line shapes are
inadequate, particularly for $T<T_c(H)$.
This is not surprising, since the predicted $d=3$ RFIM exponent
$\eta \approx 0.5$ is quite large\cite{by91}
and a large deviation from MF behavior is therefore to be expected, particularly
below $T_c(H)$, at $|t|$ much larger than in the pure $FeF_2$ case.

Since we lack a theoretically predicted line shape for data analysis,
we instead invoke the scaling form\cite{by91}
\begin{equation}
S(q) = A^{\pm}\kappa ^{\eta-2}f(q/\kappa) +B^{\pm}{A^{\pm}}^2\kappa ^{\bar{\eta}-4}g(q/\kappa)
\end{equation}
for $|q|>0$
where $\gamma = \nu (2-\eta)$ and $\bar{\gamma} = \nu (4-\bar{\eta})$.
This expression is still too complicated to use for scaling of the
data, particularly since the scaling in this unusual case
involves two perhaps independent functions.  To proceed,
we make two assumptions strongly motivated by
results of Monte Carlo simulations\cite{r95,hn98}, high temperature
expansions\cite{gaahs93}, and previous experiments\cite{bkj85} at $x=0.6$,
namely $\bar{\eta}=2\eta$ and $g(q/\kappa)=f^2(q/\kappa)$.  With these
assumptions, we have
\begin{equation}
S(q) = A^{\pm}\kappa^{\eta-2}f(q/\kappa)(1 +B^{\pm}A^{\pm}\kappa^{\eta-2}f(q/\kappa))
\end{equation}
for $|q|>0$.
Finally, we employ the TF and FB approximants for $f(q/\kappa)$ except that
we let $\phi$, $\phi ^{\prime}$, $\phi ^{\prime \prime }$,
and $\sigma $ be fitting variables.
The line shape analysis includes a small fixed constant term and one that
is linear in $q$ to account for background counts.

For $H=0$, the random-exchange Ising model case, we set $B=0$.
Folding the resolution
corrections into the scaling line shape, we fit all of the data
simultaneously over the range
$1.14 \times 10^{-4} < |t| < 10^{-2}$ for the small sample
and $1.15 \times 10^{-3} < |t| < 10^{-2}$ for the large one.
We obtain the critical
parameters $\nu = 0.70 \pm 0.02$ and $\gamma = 1.34 \pm 0.06$,
which are in good agreement with previous results using
the MF equation cited above.

The same procedure is followed for $H=7$~T, the RFIM
case, where we now let $B$ vary.
The fitted parameters are given in Table I for
$ |t| < 10 ^ {-2}$ and $|t| < 3 \times 10 ^ {-3}$.  The lower
limits for $|t|$ are the same as for $H=0$.
In Fig.\ 1 we plot, versus $q/\kappa$, the intensity data after subtracting
the background, deconvoluting with the instrumental resolution, and
dividing by $A^{\pm}\kappa^{\eta-2}(1+B^{\pm}A^{\pm}\kappa ^{\eta -2} f(q/\kappa))$.
The collapse of the data onto $f(q/\kappa)$ is excellent.  We find
no evidence of systematic deviations of the data from the scaling
function in any of the scans used in the analysis.  The collapse of the
data onto a single scaling function can only occur if the critical
exponents are chosen appropriately.  By the quality of the scaling,
we have demonstrated that by moving beyond the simple MF
line shape, it is possible to fit the RFIM
data for high magnetic concentration to obtain the critical parameters and
a good representation of the actual scattering line shape.
Note that these values, obtained for data above and
{\em below} $T_c(H)$, agree rather well with the exponents obtained
with the MF expression for this system above $T_c(H)$.
The curves in Fig.\ 1, given by the TF and FB parameters in Table 1,
represent the experimental RFIM scaling functions that can be used
to test future theoretical and simulation results.

Numerous simulation and series
expansion studies yield critical exponents for the RFIM.
The recent Monte Carlo results of Rieger\cite{r95} are 
$\nu=1.1 \pm 0.2$, $\gamma = 1.7 \pm 0.2 $, $\eta = 0.5 \pm 0.05$,
$\bar{\gamma}=3.3 \pm 0.6$, $\bar{\eta}=1.03 \pm .05$, $\beta = 0.0 \pm .005$,
and $\alpha = -0.5 \pm 0.2$.  Similarly,
Hartmann and Nowak\cite{hn98}, using exact ground state simulation techniques,
find $\nu = 1.14 \pm 0.10$, $\beta = 0.02 \pm 0.01$ and $\bar{\gamma} = 3.4 \pm 0.4$.
Note that the values for $\nu$ and $\gamma$
agree fairly well with the
results of the neutron scattering experiments (using $\gamma = \nu (2-\eta)$).
However, $\eta$ is larger than the value from the neutron study.
The value of $\beta$ has not been accurately determined experimentally.
The largest discrepancy\cite{nue98} is with $\alpha$ which appears to be close
to zero since the experiments at all concentrations exhibit
symmetric, nearly logarithmic peaks in the specific heat\cite{by91}.
This disagreement has not yet been fully explained\cite{hn98}.

The equilibrium exponents obtained in the present experiments
can be compared with previous results at lower concentrations.
Belanger, King and Jaccarino \cite{bkj85} obtained
$\nu = 1.0 \pm 0.15$, $\gamma = 1.75 \pm 0.20$, $\bar{\gamma} = 3.5 \pm 0.3$
and $\eta \approx 1/4$ using a sample of $Fe_{0.6}Zn_{0.4}F_3$ and
the MF equation.
Only data above the equilibrium boundary\cite{kjbr85} $T_{eq}(H)$,
were used.  These results, are in
good agreement with the ones obtained using
the MF line shapes in $Fe_{0.93}Zn_{0.07}F_2$.
This agreement indicates that the corrections to MF are not very significant
above $T_c(H)$ but are very large for $T<T_c(H)$.

In contrast, Feng, et al.\cite{fhbh97} obtain
$\nu=1.5 \pm 0.3$, $\gamma = 2.6 \pm 0.5 $
and $\bar{\gamma} = 5.7 \pm 1.0$ for $Fe_{0.5}Zn_{0.5}F_2$.
This discrepancy, however, may be the result of the
analysis of data which may lie below $T_{eq}(H)$,
where hysteresis prevents a study of the
equilibrium critical behavior.
This seems apparent\cite{foot} when comparing these authors' results with 
previous studies\cite{fkj91} of $Fe_{x}Zn_{1-x}F_2$.
The present study at high magnetic
concentration does away with metastability
effects entirely for the critical scattering line shape analyses.
This has the particular advantage of allowing
fits to the data on both sides of the transition and much
closer to the transition, greatly narrowing
possible interpretations of the data and yielding much more reliable
critical behavior parameters.

In summary, we have characterized the critical
scattering below $T_c(H)$ in the high magnetic concentration
crystal $Fe_{0.93}Zn_{0.07}Zn_2$.  This has not been possible
until now because of the domain formation present
in the more diluted systems previously studied.
We have shown that the line shapes are adequately described
by using simple scaling hypotheses.
The possibility still exists that the exponents obtained
in $Fe_{0.97}Zn_{0.07}F_2$ are effective
ones that may change somewhat with fields much larger than $H=7$~T. 
Future efforts will be directed to scattering
measurements to determine $\nu$, $\gamma$ and $\beta$ and
birefringence and Faraday rotation measurements to determine
the specific heat critical behavior for $H>>7$~T at high
magnetic concentration.  To avoid extinction effects,
the Bragg scattering exponent determination
requires future experiments either using neutron scattering on a
film\cite{bwshnlrl95} or using magnetic x-ray surface scattering.

We acknowledge helpful discussions with A. P. Young.  We also acknowledge
the expert technical support provided to us by R. G. Maples,
S. Moore and G. B. Taylor.  This work has been supported
by DOE Grant No. DE-FG03-87ER45324 and by ORNL, which is managed by
Lockheed Martin Energy Research Corp. for the U.S. DOE
under contract number DE-AC05-96OR22464.

\begin{figure}
\psfig{figure=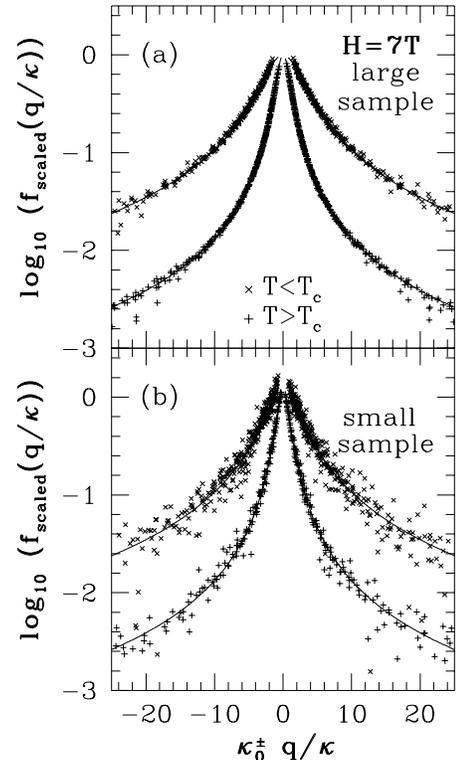,height=4.2in}
\caption{
Scaled neutron scattering data taken at different temperatures at
$H=7$ T collapsed onto the universal function $f(q/\kappa )$.
The scatter in the small sample data is larger
due to smaller number of counts obtained in the thin sample. The
fit was made for $|t|<0.01$.}
\end{figure}

\vspace{0.5in}

\begin{table}[ht!]
\begin{center}
\begin{tabular}{lcc}
parameter       &       $|t|<10^{-2}$   & $|t|<3\times10^{-3}$ \\
\hline

$T_c$ (fixed)                   &   $70.61K$          &   $70.61K$      \\
$\eta$                  &   $0.20\pm0.05$     &   $0.16\pm0.06$ \\
$\nu$                   &   $0.88\pm0.05$       &   $0.87\pm0.07$  \\
$A^+$                   &   $10.0\pm0.2$       &   $9.21\pm0.3$   \\
$A^-$                   &   $6.15\pm0.14$       &   $4.45\pm0.15$   \\
$\kappa^+_0$            &   $1.13\pm0.04$ rlu       &   $0.95\pm0.17$ rlu   \\
$\kappa^-_0$            &   $3.24\pm0.11$ rlu      &   $2.78\pm0.5$ rlu  \\
$B^+$                               &       $(4.7\pm0.1)\times10^{-5}$ &
$(3.00\pm0.13) \times10^{-5}$           \\
$B^-$                               &
$(4.0\pm0.3)\times10^{-5}$ & $(8.0\pm1.0) \times 10^{-5}$            \\
$\sigma$                &   $0.67\pm0.5$               &   $0.86\pm0.6$   \\
$\phi$                  &   $0.16\pm0.04$               &   $0.08\pm0.01$   \\
$\phi^\prime$           &   $0.39\pm0.25$               &   $0.36\pm0.3$   \\
$\phi^{\prime \prime}$  &   $0.31\pm0.25$               &   $0.26\pm0.2$   \\
$\overline{\chi^2}$     &   $3.07$               &   $2.3$           \\
No. pts.              &       $2444$                          &
$1000$                  \\
\end{tabular}
\end{center}
\caption[Fitting Parameters for the $H=7$ T Case]
{
Values found from fits at $H=7$ T.
Data for the large sample were fit for $|t| > 1.15\times10^{-3}$ and
for the small sample were fit for $|t|> 1.14\times10^{-4}$.
$\kappa $ is given in reciprocal lattice units (rlu).
}
\label{table:neut-seven}
\end{table}


\begin{references}
\bibitem{specific}  The sample is the same one used in a 
previous specific heat study\cite{sb97}.
\bibitem{note} Note that if the difference between the transition
and equilibrium temperatures were, in fact, 3.4~K, this would
essentially cover the entire critical region $|t|<0.1$ since $T_N(0)=37.6$~K,
and, hence, no equilibrium critical analysis would be possible.
\end{references}
\end{document}